\documentclass[conference]{IEEEtran}
\IEEEoverridecommandlockouts

\usepackage{cite}
\usepackage{array}
\usepackage{amsmath,amssymb,amsfonts}
\usepackage{algorithmic}
\usepackage{graphicx}
\usepackage{multirow,multicol,booktabs}
\usepackage{lipsum}
\usepackage{textcomp}
\usepackage{xcolor}
\usepackage[hyphens]{url}
\def\BibTeX{{\rm B\kern-.05em{\sc i\kern-.025em b}\kern-.08em
    T\kern-.1667em\lower.7ex\hbox{E}\kern-.125emX}}

\begin{document}

\title{Considerations in the Automatic Development of Electric Grid Restoration Plans
}

\author{\IEEEauthorblockN{Wonhyeok Jang, Hao Huang, Katherine R. Davis, Thomas J. Overbye}
\IEEEauthorblockA{\textit{Department of Electrical and Computer Engineering} \\
\textit{Texas A\&M University}\\
College Station, United States}
\{wjang777; hao\_huang; katedavis; overbye\}@tamu.edu
\vspace{-0.5cm}
}

\IEEEoverridecommandlockouts
\IEEEpubid{\makebox[\columnwidth]{978-1-7281-8192-9/21/\$31.00~\copyright2021 IEEE\hfill} \hspace{\columnsep}\makebox[\columnwidth]{ }}

\maketitle

\begin{abstract}
Power system restoration is a highly complex task that must be performed in a timely manner following a blackout. It is crucial to have the capability of developing a reliable restoration plan that can be adjusted quickly to different system conditions. This paper introduces a framework of an automated process that creates a restoration plan for a given power system. The required input includes the original system data under normal operating conditions and the status of the resources in the system that can be used for restoration. With a set of criteria to select as an option, the presented process can produce a restoration sequence in terms of which generator, load, branches, and breakers to close until a stopping criterion is met. The algorithm of the restoration process is described, and its application to a synthetic 200-bus case provides a restoration sequence from blackstart generating units to critical loads first and then to the rest of the system without violating any limits for frequency, voltage and branch loading.
\end{abstract}


\section{Introduction}
System restoration or blackstart is a procedure to restore power in the event of a partial or total shutdown of the power system. This is a highly complex decision problem with the objective of serving the load again as soon as possible without violating operating constraints. This procedure typically requires manual work by system operators, but it is time-critical. Thus, studies on restoration plans need to be conducted in advance, to aid preparation for possible blackouts. After the blackout occurred in New York City in 1977, the US Department of Energy required operating entities to develop a restoration plan for their system through NERC standards \cite{NERCStandardEOP-005},\cite{NERCStandardEOP-006}. Since then, generator operators and transmission system operators have their own restoration plans, train personnel, test their blackstart resources, and validate the plan periodically \cite{Saraf2009a,PJM2020,IESO2017,CaliforniaISO2017}. Also, both the academia and the industry have put significant effort into this topic, and some of the results have been published as a compendium of papers in 1993, 2000, and 2014 \cite{PowerSystemRestorationWorkingGroup1993, Adibi2000,RestorationDynamicsTaskForce2014}.

Traditionally, a system restoration process is categorized into two approaches: a top-down method and a bottom-up method. The top-down approach utilizes neighboring power systems to energize the blacked-out system via tie lines, while the bottom-up approach begins the process with the pre-selected generating units with the self-start capability within the operation territory. Recently, with the deployment of microgrids and distributed energy resources (DERs), various restoring strategies have been proposed, especially for distribution systems. For instance, researchers in \cite{Song2013,Che2014,Byeon2020} present operating schemes of a distribution system using DERs for service restoration. Also, other studies show how to optimally invest and schedule microgrids for different contingencies in the systems \cite{Khodaei2013,Gholami2016,Wang2018}. For transmission networks, deployment of Wide Area Monitoring Systems (WAMS) utilizing electronic switching devices and advanced monitoring devices such as phasor measurement units contribute a more reliable system restoration process by making early warning systems possible, helping the restoration process, and improving post-event analysis \cite{Novosel2008a}. As the alternative for mechanical circuit breakers, high-power high-speed thyristors can serve as a substitute without depleting the stored energy in substation, and thus WAMS devices can operate throughout the blackout period aiding the restoration process \cite{Adibi2010}. Optimization techniques were also utilized to tackle restoration ordering/routing problems to minimize the size of the blackout \cite{Mak2014,VanHentenryck2015,Bienstock2007}.
Even with numerous studies on the topic, blackstart is still a complex task, and a single strategy is not sufficient to account for the wide range of potential system conditions. For example, a different restoration process may be required for the same system based on the status of available resources after a blackout. Restoration planning studies are primarily performed offline using simulations due to limited online implementation and most operating entities already have their own restoration plans for a total blackout. However, what if the set of available resources assumed in the previously developed restoration plan is different when a blackout actually happens? What if an extreme disaster destroyed some of the key resources, and a big part of the pre-established restoration plan is no longer possible to implement? In such cases, following the existing plan would not be optimal, especially if operators do not have the capability to redesign the restoration plan in a timely manner. Therefore, the authors present a framework of an automated restoration process for a given power system and its condition. The goal is to produce a reliable restoration procedure for a system after a blackout so that the pre-specified portion of the load is back to online without violating the requirement of frequency, voltage and branch limits based on the built-in options for users to choose.

\section{Overview Of The Restoration Process}\label{Sec:ProcessOverview}
The first step in a restoration process is to evaluate the system status. A fast but reliable blackstart is only possible with accurate status determination. Assessment of available resources such as which generators can be dispatched, which lines can be energized, etc. must be performed first for the system operators to determine the next step \cite{podmore2010smart}. The presented automatic restoration process also requires such input from users that indicates which generators, lines, transformers, etc. are available. The work presented in the paper used a PowerWorld Simulator case as an input and the final product is the same case file with a complete restoration sequence added under Transient Stability, which can be exported to different formats. In addition, blackstart units (BSUs) and critical loads (CLs) have to be specified by users. Parallel restoration of subareas in a blacked-out system is often desired when possible and the presented automated process also supports that. One can specify subarea information including BSUs and CLs if they want to restore those subareas in parallel at the same time. Below are a few criteria for users to choose, based on their preference for restoration.

\subsection{Criteria for Blackstart Process}
Each system has its own topology, characteristics, and blackstart resources, and critical loads, thus a single restoration approach may not be appropriate for different systems. Even for the same system, there should be different procedures based on the availability of blackstart resources and the transmission network after a blackout. In addition, each utility may have different priorities when restoring their system. Possible priorities include total cost, amount of system load to be restored, or desired restoration time \cite{Saraf2009a}. Thus, the process offers a few criteria to choose as an option to accommodate various preferences as described below.

\subsubsection{Criterion 1 -- whether to pick up a generator/load next}
A generator is picked up if a portion of the total available online generator MW output is smaller than the load MW to pick up next. A load is picked up otherwise. Users can set the portion between 0\% and 100\% depending on system characteristics and preference. Here, the total available online generator MW output is the difference between the sum of maximum MW output and the sum of the current MW output of all the online generators. This criterion controls the timing of generator pickups. With a smaller value, more generators will be picked up earlier securing enough generation for the next load to pick up, but with the price of serving load later.

\subsubsection{Criterion 2 -- which generator to pick up}
Users can choose which generator to pick up next based on maximum MW output, minimum MW output, start-up time, the distance from the online area, or the time to dispatch a crew.

\subsubsection{Criterion 3 -- which load to pick up}
Users can choose which load to pick up next based on maximum MW, minimum MW, the time to dispatch crews, etc.

\subsubsection{Criterion 4 -- whether to stop the restoration process}
It is possible to set how much percentage of the total load MW in the original case should be online for stopping the restoration process.

\subsection{Overall Restoration Process}
\begin{figure}[tb]
\vspace{-0.5cm}
	\centering
	\includegraphics[width=0.45\textwidth]{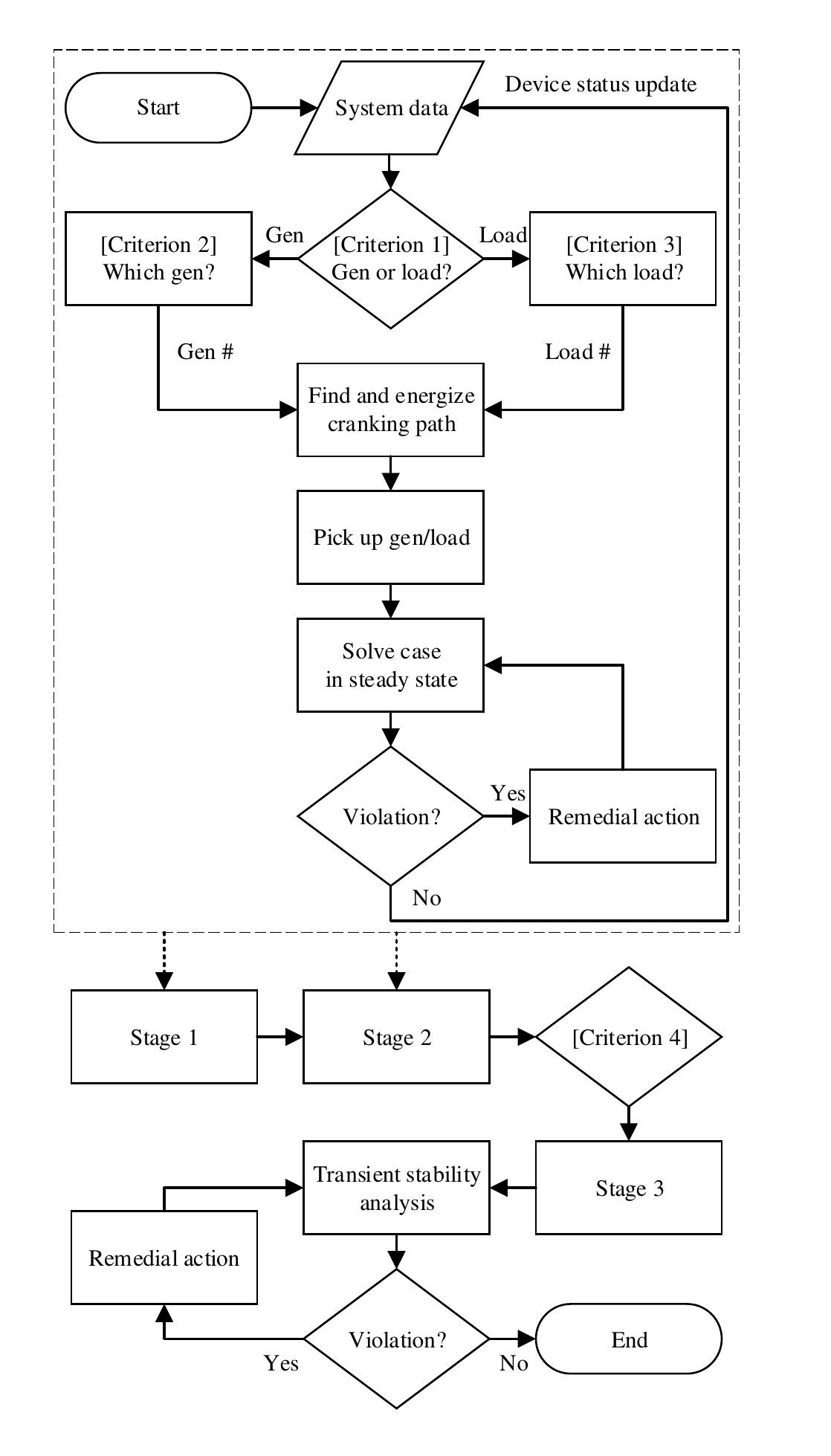}
	\vspace{-10pt}
	\caption{Proposed restoration process}
	\label{fig:Flowchart}
	\vspace{-0.5cm}
\end{figure}

The overall restoration process illustrated in Fig. \ref{fig:Flowchart} has three stages as described below.
\begin{itemize}
\item {Stage 1: restoration of BSUs and CLs}\\
This stage starts with picking up BSUs and CLs and finishes when all the CLs are online. This may include picking up some non-blackstart units (NBSUs) when available BSUs are not large enough to supply all the CLs. Thus, \textit{Criterion 2} and \textit{3}  gives higher priority to BSUs and CLs over NBSUs and non-critical loads (NCLs) in this stage, respectively.
\item {Stage 2: restoration of NBSUs and NCLs}\\
This stage restores the remaining offline part of the system. Cranking paths between the online area to NBSUs and NCLs are identified and energized. Majority of the NBSUs and NCLS will be picked up in this stage until \textit{Criterion 4} is met.
\item {Stage 3: synchronization}\\
In this stage, tie lines between all the subareas restored in Stage 2 are energized. This includes restoring the subareas without BSUs by energizing cranking paths from the neighboring subareas. This stage is not required if a user does not want parallel subarea restoration.
\end{itemize}

\subsection{Procedure for Generator/Load Pickup}
The detailed procedure for picking up generators and loads based on \textit{Criterion 1-3} is illustrated in a dashed box in Fig. \ref{fig:Flowchart}. It should be noted that the entire module of \textit{Criterion 1-3} is subject to customization based on the objective of the restoration with different priorities. This procedure is applied to Stage 1-2 whenever picking up a generator or a load. This utilizes the available generator data, load data, each substation’s bus topology, network information with available transmission lines and admittance matrix to determine the cranking path from the online area to the next gen/load to pick up. Once the generator or load to be picked up next is determined by \textit{Criterion 1-3}, the cranking path to the device needs to be determined. One simple but effective way to pick up a device is through the path with the least impedance from the online area which would result in less losses.
To calculate the minimum electrical distance between two buses, the shortest path between the online area and the next gen/load to pick up is calculated using Dijkstra’s algorithm \cite{Dijkstra1959}. Edges in the network graph are weighted by the absolute value of the impedance from the network admittance matrix, $Y$. The weight of the edge between bus \textit{i} and \textit{j} is,

\begin{equation}
C_{ij} = \left| \frac{1}{Y_{ij}} \right|
\end{equation}

Once the next generator or load to pick up is determined, its bus becomes the source node and all the online buses in the subarea become the target nodes to calculate their path length from the source node. The path with the minimum total weight is selected as the cranking path to be energized.

Running a transient stability (TS) simulation whenever picking up a device could be time consuming. To reduce the overall simulation time, the case is solved in steady state when a device is picked up during Stage 1-2 as shown in the dashed box in Fig. \ref{fig:Flowchart}. When there are any violations of monitored limits, remedial actions are performed automatically. When the restoration process is over after Stage 3, TS simulations are performed to see if further remedial actions are needed to mitigate limit violations. Finally, a complete list of sequences of closing all the devices, down to the breaker level is produced. Users can perform studies with the outcome and make further adjustments if necessary.

\subsection{Limit Monitoring}
As the process restores a system, it is possible to have violations that could jeopardize the secure and stable restoration process. We have determined three parameters to monitor and provide feedback to mitigate any possible violation during the entire restoration process: voltage, frequency, and branch loading. All three parameters are monitored when the case is solved after a device is picked up in both steady state and in transient state. When they violate the limits, remedial actions are employed. Since the initial stage of the restoration process tends to have more fluctuation in voltage and frequency, their monitoring limits are wider than those in typical normal operating conditions.

\begin{itemize}
\item {Voltage}\\
Overvoltage may occur due to the line charging when energizing high voltage transmission lines and due to harmonic resonance during switching operations. Also, undervoltage may occur when large reactive loads are picked up. There are various ways of providing voltage control, but here voltage fluctuations are mitigated by changing the generators’ exciter setpoints and/or closing/opening available shunt capacitors if possible. It should be noted that discrete shunt capacitors may cause overvoltage, thus extra care is required when closing them. Additionally, loads are picked up incrementally for generators to have enough time to respond. The amount of load MW increment can be set up by users.
\item {Frequency}\\
Under-frequency is more likely to happen than over-frequency during the restoration process. To tackle this, first, the MW output is increased for the generators close to the buses with low frequency. If there are no available generators nearby or all the nearby generators are running at their maximum output, the load shedding is performed. Also, load pickup increment setting helps the frequency control. It should be noted that automatic generation control is set to be off for all the generators during the restoration process.
\item {Branch loading}\\
Branch overloading is mitigated  by closing a set of offline branches that can reroute some of the power flow from the branches with a loading percentage higher than the pre-specified limit. To perform this efficiently, line close distribution factor (LCDF) is calculated for the pair of buses of the most overloaded branch and the offline branch with the maximum LCDF is closed.
\end{itemize}

\section{Test Case}
\subsection{Synthetic 200-bus Case}
\begin{figure}[htb]
	\centering
	\includegraphics[width=0.35\textwidth]{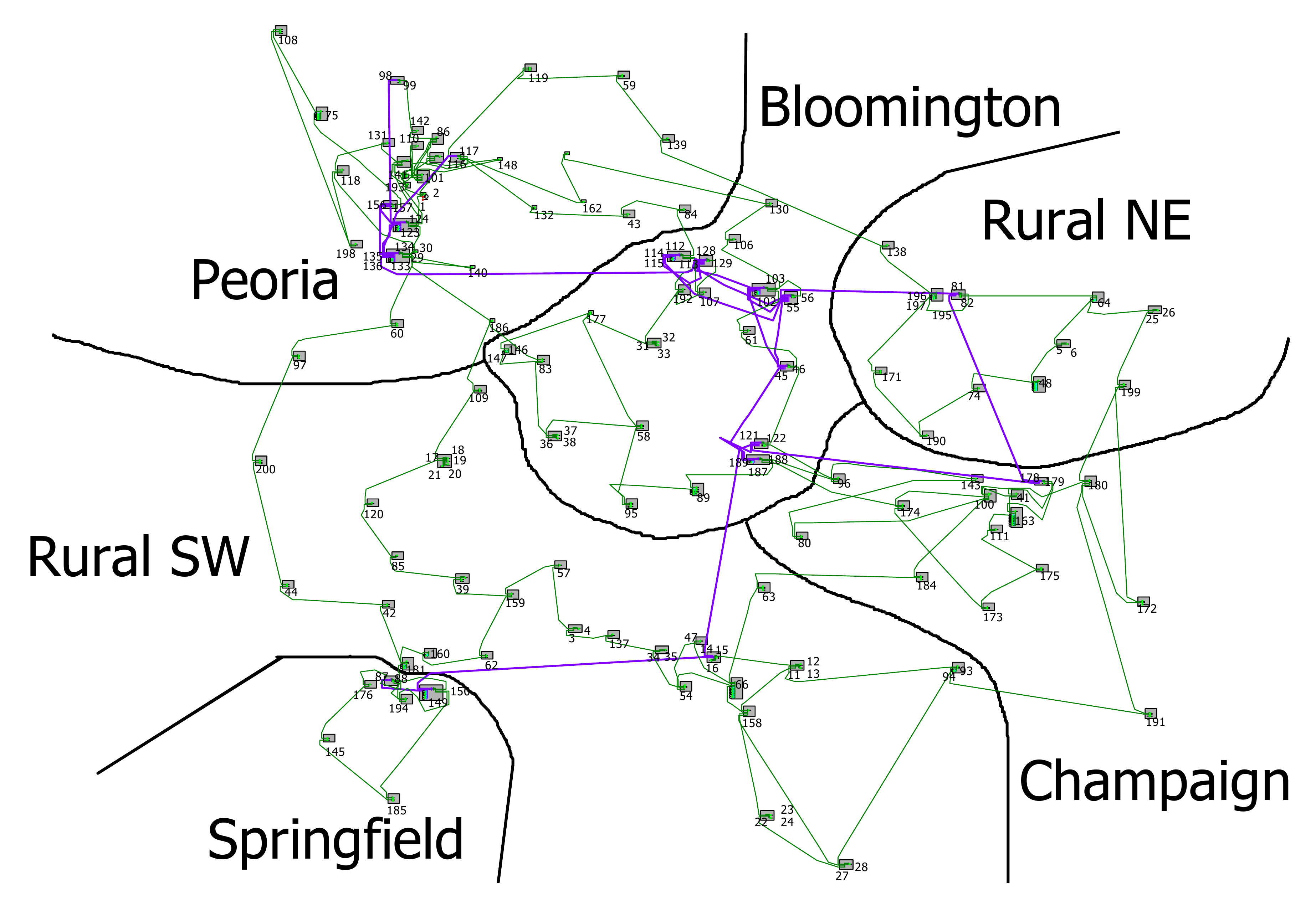}
	\vspace{-0.1cm}
	\caption{Synthetic 200-bus central Illinois case}
	\label{fig:200buscasemap}
	\vspace{-0.5cm}
\end{figure}

Due to the confidentiality issues with real power system models and their data, this paper utilizes a synthetic 200-bus system, which is divided into six zones, as shown in Fig. \ref{fig:200buscasemap} \cite{TAMUrepository}. This synthetic grid case was built on the geographic footprint of central Illinois and is statistically similar to the actual power system of the area in terms of numbers and locations of loads, generations and transmission lines, but the system is fictitious, and thus does not have any security concerns. The algorithms, statistical data, dynamic models and the metrics used in creating this realistic yet synthetic case are presented in \cite{Birchfield2017,Birchfield2017a,Xu2018}. The data for each of them is presented in Table \ref{table:CaseData}.

\begin{table*}[t]
\centering
    \caption{Data of the synthetic 200-bus central Illinois case divided into six zones}
    \vspace{-0.1cm}
    \begin{tabular}{>{\centering\arraybackslash}m{1.5cm} >{\centering\arraybackslash}m{0.6cm} >{\centering\arraybackslash}m{0.6cm} >{\centering\arraybackslash}m{0.6cm} >{\centering\arraybackslash}m{0.6cm} >{\centering\arraybackslash}m{0.6cm} >{\centering\arraybackslash}m{0.6cm} >{\centering\arraybackslash}m{0.6cm} >{\centering\arraybackslash}m{0.6cm} >{\centering\arraybackslash}m{0.6cm} >{\centering\arraybackslash}m{0.6cm} >{\centering\arraybackslash}m{0.6cm} >{\centering\arraybackslash}m{0.8cm} >{\centering\arraybackslash}m{0.6cm} >{\centering\arraybackslash}m{0.6cm} >{\centering\arraybackslash}m{0.6cm}}
    \toprule
        \multirow{3}{*}{Zone} & \multicolumn{5}{c}{Generators} & \multicolumn{6}{c}{Loads} & \multicolumn{2}{c}{Branches} & \multicolumn{2}{c}{Switched shunts}\\ \cmidrule(lr){2-6}\cmidrule(lr){7-12}\cmidrule(lr){13-14}\cmidrule(lr){15-16}
         & \multicolumn{3}{c}{Total} & \multicolumn{2}{c}{BSU} & \multicolumn{3}{c}{Total} & \multicolumn{3}{c}{BSU} & \multicolumn{2}{c}{Num.} & \multirow{2}{*}{Num.} & \multirow{2}{*}{Mvar}\\ \cmidrule(lr){2-4}\cmidrule(lr){5-6}\cmidrule(lr){7-9}\cmidrule(lr){10-12}\cmidrule(lr){13-14}
         & Num. & Max. MW & Max. Mvar & Max. MW & Max. Mvar & Num. & MW & Mvar & Num. & MW & Mvar & Within Zone & Tie-line \\
         \midrule
         Peoria & 9 & 871 & 225 & 18 & 9 & 41 & 663 & 189 & 2 & 42 & 12 & 63 & 7 & 0 & 0\\
         Springfield & 5 & 314 & 32 & 77 & 36 & 8 & 202 & 58 & 2 & 30 & 9 & 15 & 2 & 1 & 50 \\
         Rural SW & 11 & 276 & 53 & 139 & 71 & 42 & 361 & 103 & 2 & 35 & 10 & 53 & 7 & 1 & 30\\
         Champaign & 7 & 6 & 10 & 26 & 13 & 25 & 367 & 105 & 1 & 42 & 12 & 23 & 9 & 1 & 80 \\
         Rural NE & 8 & 51 & 31 & 67 & 34 & 12 & 93 & 27 & 2 & 36 & 3 & 22 & 4 & 0 & 0 \\
         Bloomington & 9 & 685 & 145 & 6 & 3 & 32 & 492 & 140 & 0 & 0 & 0 & 51 & 12 & 1 & 30 \\
         \bottomrule
    \end{tabular}
    \label{table:CaseData}
    \vspace{-0.5cm}
\end{table*}

Each zone in the test case has its own BSUs and CLs, thus they are restored independently of each other in parallel at the same time to speed up the overall process. In reality, each island would be restored as soon as it is ready regardless of the status of other islands, thus reducing the time for consumers not being served. In the test case, all the zones have more generation capacity than load except Champaign. For the subareas where load is greater than generation, only part of the load will be picked up to the available generation capacity in the subarea until Stage 2, and the rest of the load will be served in Stage 3 when the subarea is synchronized with its neighbors. The number of branches in Table \ref{table:CaseData} includes the transmission lines and transformers but excludes branches with zero impedance which account for the lines between busbars and breakers in substations.

\subsection{Node Breaker Topology}
The original 200-bus case has a bus-branch configuration. For developing a restoration process, it is important to know which breakers to close in identifying cranking paths. Thus, the test case was converted to a node-breaker configuration, and all the substation topologies were converted based on the highest rated bus voltage as presented in Table \ref{table:SubTopoloty}. The status of all the breakers in the blacked out area is assumed to be open following the “all-open” strategy, to simplify determination of network status \cite{Adibi1994}, but designation of unavailable devices is also possible in the case file as the input.

\vspace{-0.2cm}
\begin{table}[!htb]
    \centering
    \caption{Substation topologies based on the highest voltage level}
    \vspace{-0.1cm}
    \begin{tabular}{c c}
        \toprule
        Highest rated voltage & Topology\\ \midrule
        $230 kV $ & Double-bus-double-breaker\\
        $115 kV $ & Breaker-and-a-half\\
        $13.8 kV $ & Single bus\\
        \bottomrule
    \end{tabular}
    \label{table:SubTopoloty}
    \vspace{-0.5cm}
\end{table}

\section{Simulation Conditions and Results}\label{Sec:Sim}
\subsection{Simulation Settings}
Table \ref{table:SimSetting} shows simulation settings used specifically for the synthetic 200-bus test case. It should be noted that different settings may result in a different total restoration time and system configuration. 5\% for \textit{Criterion 1} indicates that most of the generators are picked up in the earlier phase of the restoration process for stable operations even though loads are picked up a bit later. Larger generators are picked up earlier, and smaller loads are picked up earlier with 20 MW of increment if it is greater than 20 MW. Six zones are restored independent of each other in parallel at the same time and tie lines are energized when all the zones are stable at the end of the process. The time interval is placed after picking up a device. When picking up loads, it takes a longer time for the system to stabilize than other devices such as generators, branches, and shunts, thus having a longer time until the next pickup. These time intervals are set based on the preliminary simulations and are case dependent. When users utilize practical time periods for \textit{Criterion 2-3} such as generator ramp time, crew dispatch time, etc., these times intervals can reflect those time scales. 

\vspace{-0.2cm}
\begin{table}[hbt]
    \centering
    \caption{Simulation settings for the 200-bus case}
    \vspace{-0.1cm}
    {\renewcommand{\arraystretch}{1.0}
    \begin{tabular}{ >{\centering\arraybackslash}m{1.3cm} m{5.1cm} m{1.3cm} } 
        \toprule
        \multicolumn{1}{c}{Option} & \multicolumn{1}{c}{Description} & \multicolumn{1}{c}{Setting}\\ \midrule
        \textit{Criterion 1} & Portion of the total available online generator MW output with respect to the load MW to pick up next & 5$\%$\\ \cmidrule(lr){1-3}
        \textit{Criterion 2} & Which generator to pick up next & Max. MW\\ \cmidrule(lr){1-3}
        \textit{Criterion 3} & Which load to pick up next & Min. MW\\ \cmidrule(lr){1-3}
        \textit{Criterion 4} & Portion of online load MW with respect to the total load MW & 80$\%$ \\ \cmidrule(lr){1-3}
        \textit{Load\_inc} & MW increment when picking up a load & 20 MW \\ \cmidrule(lr){1-3}
        \textit{Parallel} & Parallel restoration for subareas & Yes \\ \cmidrule(lr){1-3}
        \textit{Gen Vref} & Voltage setpoint when picking up a generator & 1.04 p.u. \\ \cmidrule(lr){1-3}
        \textit{T\_Load} & Time interval after picking up a load & 20 s \\ \cmidrule(lr){1-3}
        \textit{T\_Event} & Time interval after picking up a generator/branch/shunt & 10 s \\
        \bottomrule
    \end{tabular}}
    \label{table:SimSetting}
\end{table}

Table \ref{table:Limits} displays the limit values of the three monitored parameters used for the test case. Both bus voltages and frequencies have time varying lower and upper limits for the remedial actions to be employed. On the other hand, branch loadings have just an upper limit and as soon as any branch is loaded more than the specified limit, the mitigation measures are taken as described in Section \ref{Sec:ProcessOverview}. Theses limit values can be adjusted based on system characteristics and user preferences.

\vspace{-0.1cm}
\begin{table}[!htb]
\vspace{-0.1cm}
    \centering
    \caption{Limit values of monitored parameters for the 200-bus case}
    \vspace{-0.1cm}
    {\renewcommand{\arraystretch}{0.9}
    \begin{tabular}{>{\centering\arraybackslash}m{1.2cm} >{\centering\arraybackslash}m{1.2cm} >{\centering\arraybackslash}m{3.6cm} >{\centering\arraybackslash}m{1.1cm}} 
        \toprule
        Parameter & Type & Limit value & Violation duration\\ \midrule
        \multirow{2}{*}{\parbox{1.2cm}{\centering Bus voltage}} & Instant & $0.8 < V_{pu} < 2.0$ & 0 s\\ \cmidrule(lr){2-4}
        & Sustained & $0.95 < V_{pu} < 1.10$ & 10 s \\ \cmidrule(lr){1-4}
        \multirow{2}{*}{\parbox{1.2cm}{\centering Bus frequency}} & Instant & $59\ Hz < Freq. < 61\ Hz$ & 0 s\\\cmidrule(lr){2-4}
         & Sustained & $59.6\ Hz < Freq. < 60.4\ Hz$ & 10 s \\ \cmidrule(lr){1-4}
        Branch loading & Instant & $ BR_{loading} < 90 \%$ & 0 s\\
        \bottomrule
    \end{tabular}}
    \label{table:Limits}
    \vspace{-0.2cm}
\end{table}

The renewable power sources are typically assumed to be unavailable during restoration, as they do not provide stable generation. Thus, all the wind generators were set to be disconnected from the grid throughout the process.

\subsection{Simulation Results}
\begin{figure}[!htb]
	\centering
	\includegraphics[width=0.41\textwidth]{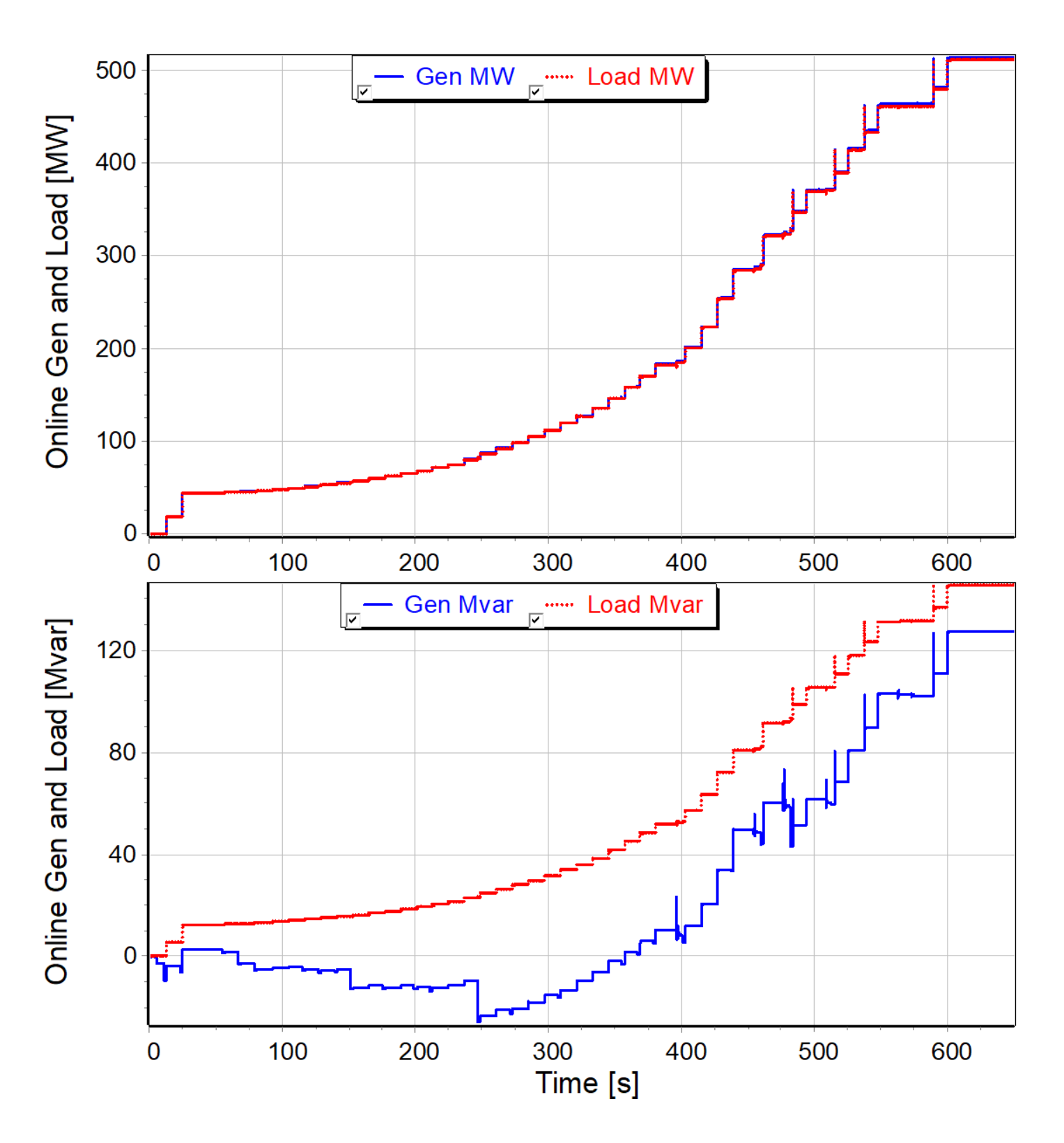}
	\vspace{-0.5cm}
	\caption{Total online generation and load during restoration in Peoria}
	\label{fig:GenLoadMWMvar}
\end{figure}

Fig. \ref{fig:GenLoadMWMvar} shows the total online generation and the load in Peoria both in MW and Mvar during restoration as an example. Picking up a large load could make a system fluctuate especially when there are not much of online generation during restoration. Preliminary simulations indicated that 20 MW is an appropriate threshold for the test case and thus they are picked up incrementally and generation closely follows the load profile until \textit{Criterion 4} is met around 600 s.
\vspace{-0.5cm}

In addition, the generators consume reactive power until 350 s due to the excessive reactive power provided by lightly loaded lines in the system. This phenomenon disappears as more loads are picked up along the process. Adjusting \textit{Criterion 1} affects on this as more or fewer lines are energized to pick up generators before some of the loads are being online.

\begin{figure}[htb]
	\centering
	\includegraphics[width=0.41\textwidth]{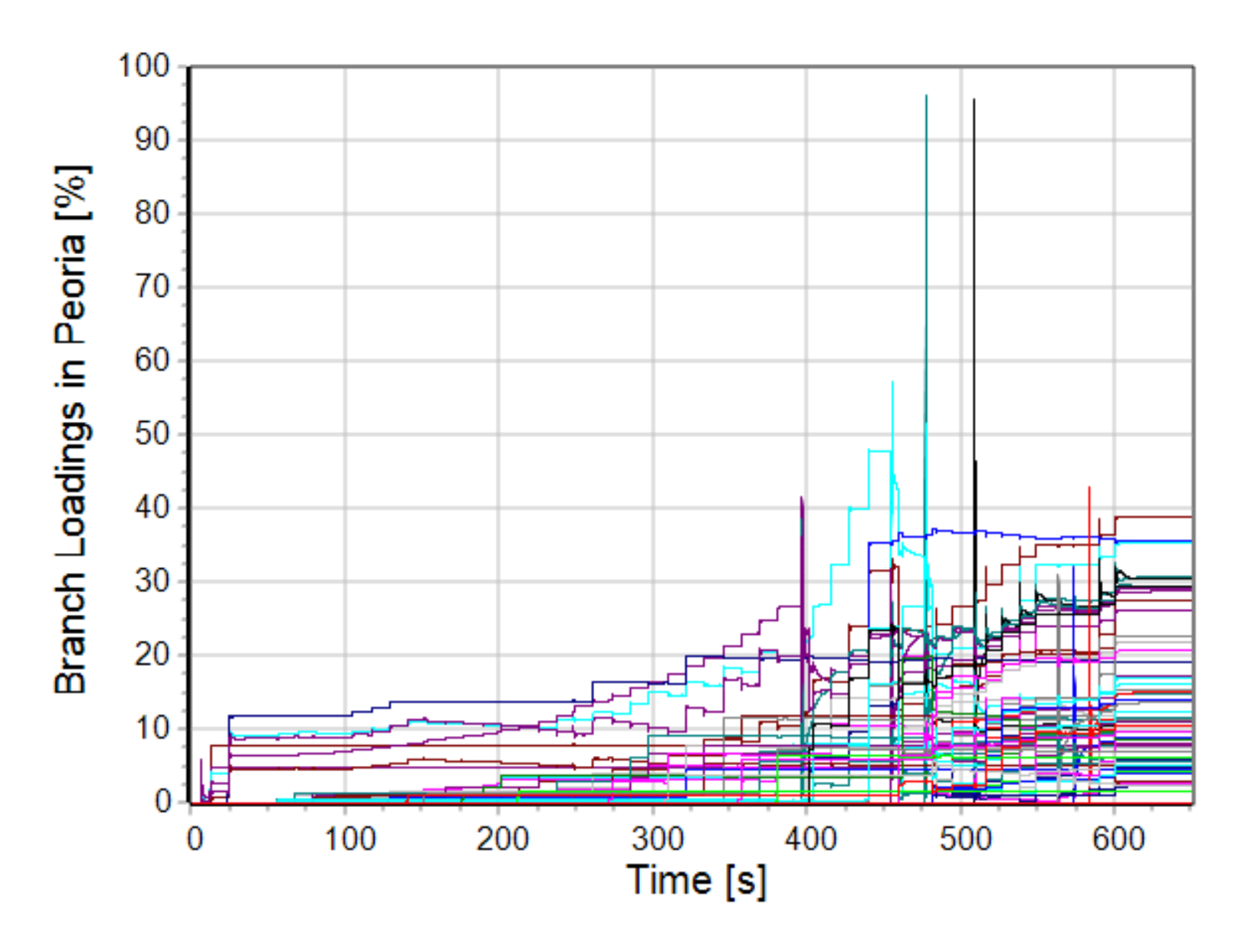}
	\vspace{-0.4cm}
	\caption{Branch loading during restoration in Peoria}
	\label{fig:BRloading}
	\vspace{-0.2cm}
\end{figure}

Fig. \ref{fig:BRloading} shows the branch loading in percentage in Peoria. Some of the lines in the zone are heavily loaded near 480 s and 510 s. As branch loadings go over the threshold (90\%), they are mitigated by closing the nearby branches as a remedial action mentioned in Section \ref{Sec:ProcessOverview}.

\vspace{-0.5cm}

\begin{figure}[htb]
	\centering
	\includegraphics[width=0.41\textwidth]{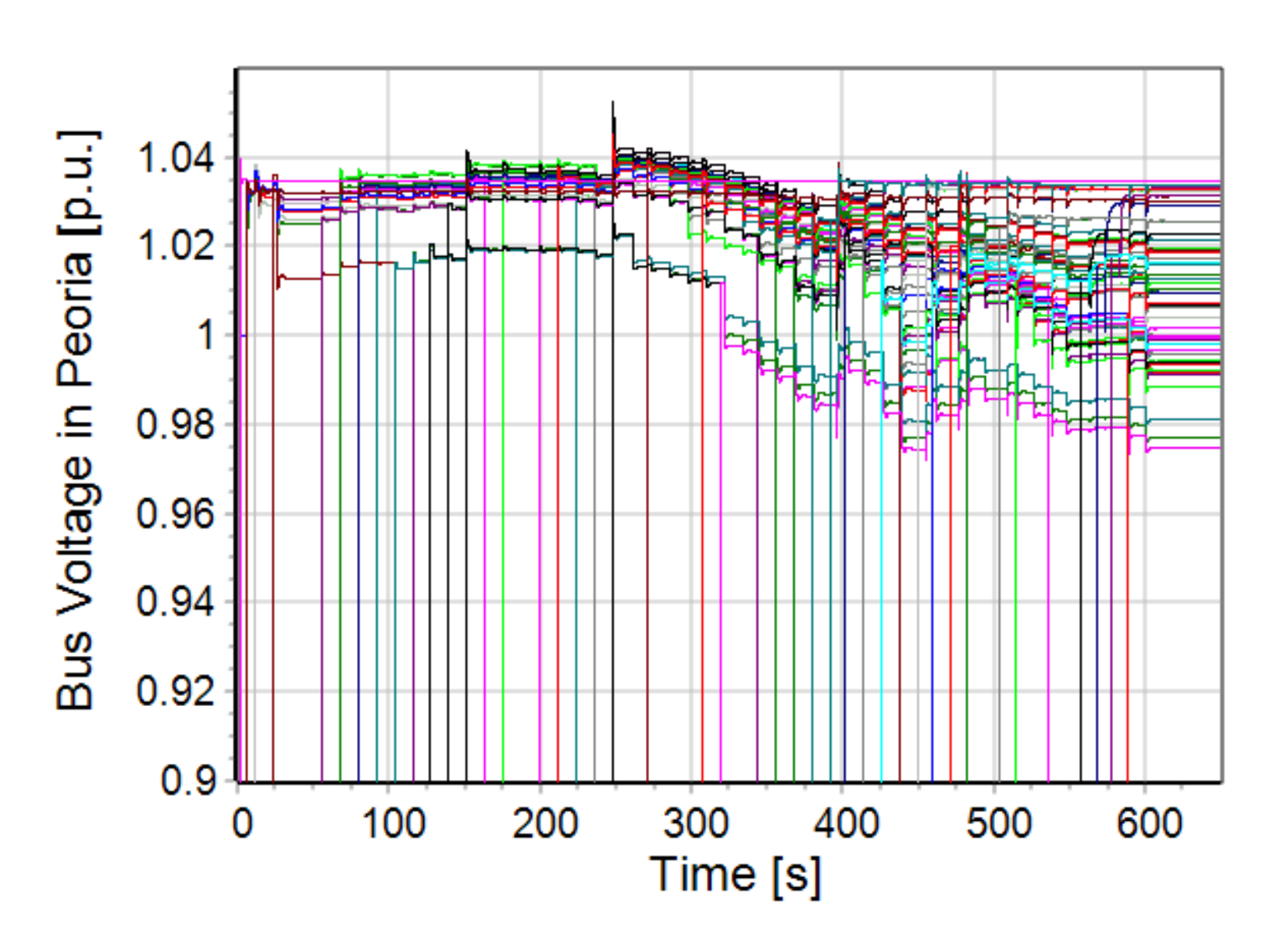}
	\vspace{-0.4cm}
	\caption{Bus voltage during restoration in Peoria}
	\label{fig:Vpu}
	\vspace{-0.2cm}
\end{figure}

Bus voltages during the restoration process in Peoria are shown in Fig. \ref{fig:Vpu}. Some of the buses have their voltages lower than 0.98 p.u., but any remedial actions were not performed as they were over the lower limit of 0.95 p.u. specified in Section \ref{Sec:Sim}. All the vertical lines in the figure were drawn when buses are energized resulting the p.u. bus voltage to jump from zero. The limits for voltage monitoring can be narrower for the price of a longer overall restoration time. Even though the test case has four switched shunts available as shown in Table \ref{table:CaseData}, they do not have a discrete output control. Extra simulations showed that inserting them causes overvoltage, thus none of them were utilized during the restoration.

\vspace{-0.5cm}

\begin{figure}[htb]
	\centering
	\includegraphics[width=0.49\textwidth]{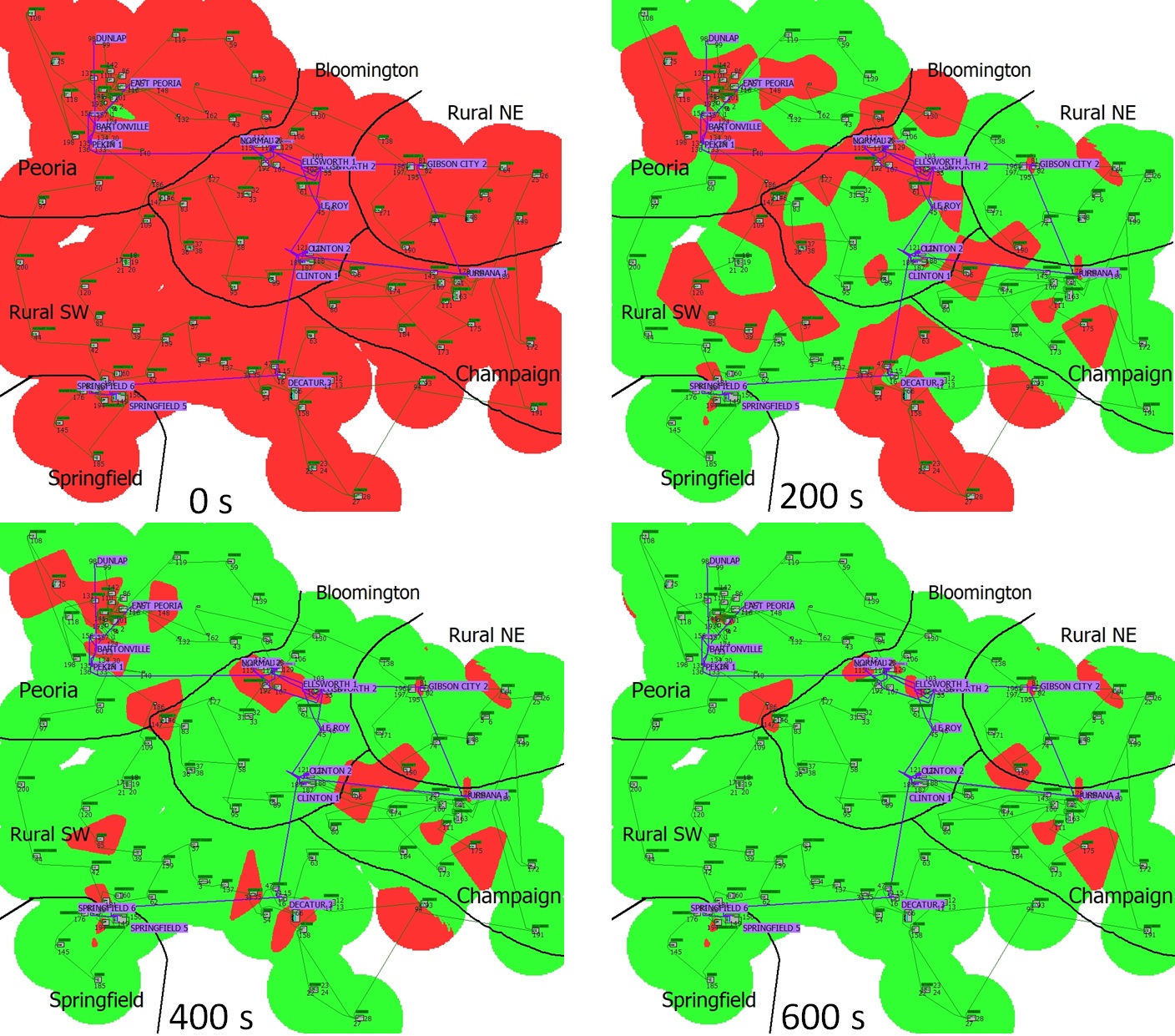}
	\caption{Bus frequency during restoration in 200-bus case}
	\vspace{-0.5cm}
	\label{fig:Freq}
\end{figure}

Fig. \ref{fig:Freq} illustrates the entire system as each zone is being restored in parallel with a color contour used to visualize the bus frequency \cite{Weber2000}. The green indicates the nominal frequency of 60 Hz and the red is for 0 Hz. Bus frequencies were monitored throughout the process and generator outputs were controlled to maintain them within the specified limits. Since the stopping criterion for restoration was set to 80\% of the total load, there are still a few offline buses in red at the end of the restoration.

\section{Conclusions}
The framework of developing an automated power system restoration plan is introduced in the paper. The detailed restoration process is presented with a set of criteria and options for users to choose from. The current built-in options provide users general objectives to be desired in a typical restoration process. The proposed automated process produces a complete sequence of devices to close from a completely blacked-out system while making sure each step of device energization has no limit violations in both steady state and transient state. Since this is the early stage of developing the automatic power system restoration plan, there are more to be added to make it more comprehensive in the future. One example is that a set of generators and loads may be grouped as critical resources and be picked up at the same time instead of one by one within each subarea to further reduce the restoration time. In addition, different existing optimal restoration solutions will be incorporated to meet various preferences. When this process has more features and improvements, it will benefit people in the field not only in developing restoration procedures for their system but also for comparison and enhancement of their own plans with the outcome of this automated process.

\section*{Acknowledgment}
The work presented in this paper was partially supported by the US Department of Energy (DOE) Advanced Research Projects Agency-Energy (ARPA-E), DOE Cybersecurity for Energy Delivery Systems program under award DE-OE0000895 and the National Science Foundation under Grant 1916142.

\bibliographystyle{IEEEtran}
\bibliography{Naps2020_restoration.bib}

\end{document}